# On the significance in signal search through the "sliding window" algorithm

Gioacchino Ranucci
*Istituto Nazionale di Fisica Nucleare*
*Via Celoria 16 - 20133 Milano*
*e-mail: gioacchino.ranucci@mi.infn.it*

The experimental issue of the search for new particles of unknown mass poses the challenge of exploring a wide interval to look for the usual signatures represented by excess of events above the background. A side effect of such a broad range quest is that the traditional significance calculations valid for signals of known location are no more applicable when such an information is missing. In this note the specific signal search approach via observation windows sliding over the range of interest is considered; in the assumptions of known background and of fixed width of the exploring windows the statistical implications of such a search scheme are described, with special emphasis on the correct significance assessment for a claimed discovery.

# 1. Introduction

While searching for a signal against a noise background the significance level of a claimed detection is different depending upon whether the location of the signal is known or not. In the former case, and assuming the background known, the standard situation of comparing in the region of interest the detected number of counts with the expected background distribution (typically the Poisson distribution for counting experiments) holds; in the latter the a-priori unknown location of the signal forces the adoption of different search strategies which alter this simple picture. The simplest of such strategies is the investigation of the whole region to be scanned for the signal via a set of windows (the so called sliding windows) of specified width: signal detection is thus claimed if in at least one of these windows a number of events is counted to be above the detection threshold. The related issue to be addressed in this strategy is how to fix the threshold so to ensure the correct desired confidence level for a detection claim.

Such a matter has been recently considered in [1], [2], [3], [4] and [5]. In particular in [4], in the framework of studies done in view of the new particles search that will be carried out at the LHC, it has been extensively demonstrated via Monte Carlo how the statistics nature of the multiple windows search process is altered with respect to the simple single window search; specifically it has been shown that the use of the detection thresholds which in the single search correspond to the usual confidence levels would lead in the multiple search strategy to overestimate the signal discovery capability.

Purpose of this short note is to establish the origin of the problem of correctly assessing the desired detection confidence levels in the sliding window algorithm in a different way, through the use of a demonstration based on the mathematical formulation of the occurrence probabilities associated with the multiple decision process. Some Monte Carlo calculations to validate the derived model are presented, too.

# 2. Occurrence probabilities for the maximum detected count in a multiple windows search process

Let's assume that we scan a region of interest through a number W of windows of equal width, looking for an excess of events as signature of a new particle, and that the background over each window is common and equal to B. The situation we want to study is that in which, the signal being not present, we claim however a false detection because of noise fluctuations that, at least in one of the scanned windows, produce a number of counts above the detection threshold. In order to compute the probability of such an occurrence we need to evaluate the probability distribution of the highest number of counts detected over the whole set of searching windows, since it is on this highest count that we set the detection threshold in order to declare a discovery. If the W windows are not overlapping such a calculation can be done exactly.

Let's assume that in each window the count rate is dictated by the usual Poisson distribution

$$P(n, B) = \frac{e^{-B} B^n}{n!} \qquad (1).$$

Our goal is to derive from this original distribution the distribution of the largest detected count N over the whole set of searched windows. As usual in probability demonstrations, the distribution of N can be inferred by considering all the configurations that produce each specific realization of such a random variable. In particular, a given value N is obtained when in all the windows but one the counts are less than N, while in the residual window are exactly N; the probability associated to such a configuration is clearly

$$W\left(\sum_{n=0}^{N-1}\frac{e^{-B}B^n}{n!}\right)^{W-1}\frac{e^{-B}B^N}{N!} \qquad (2)$$

where the factor W takes into account the number of combinations for choosing one window out of the total number W.

The same maximum count value N is also obtained when in all the windows but two the counts are less than N, while in the two residual windows they are exactly N; the probability associated to such a configuration is

$$\binom{W}{2}\left(\sum_{n=0}^{N-1}\frac{e^{-B}B^n}{n!}\right)^{W-2}\left(\frac{e^{-B}B^N}{N!}\right)^2 \qquad (3)$$

where the factor $\binom{W}{2}$ takes into account the number of combinations for choosing two windows out of W. Generalizing such a consideration, the $k_{th}$ configuration producing the same maximum count N is associated to the probability

$$\binom{W}{k}\left(\sum_{n=0}^{N-1}\frac{e^{-B}B^n}{n!}\right)^{W-k}\left(\frac{e^{-B}B^N}{N!}\right)^k \qquad (4)$$

and hence the total probability for N is given by

$$P_{\max}(N)=\sum_{k=1}^{W}\binom{W}{k}\left(\sum_{n=0}^{N-1}\frac{e^{-B}B^n}{n!}\right)^{W-k}\left(\frac{e^{-B}B^N}{N!}\right)^k \qquad (5).$$

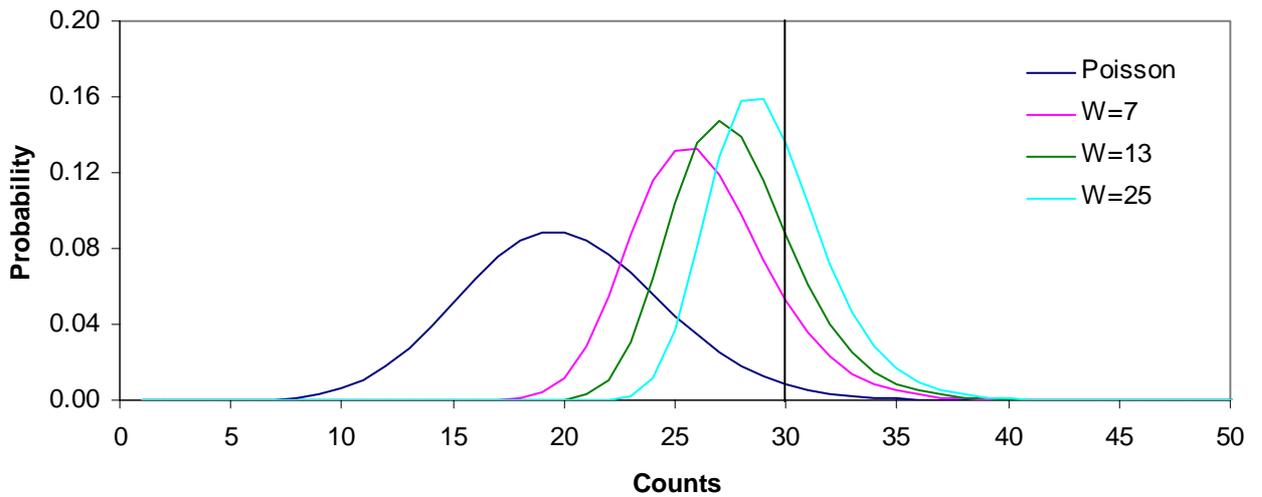

*Fig. 1. Original Poisson distribution and modified noise distributions for different numbers of exploring windows related to the toy model described in the text. The vertical line marks the $2\sigma$ threshold on the Poisson curve*

The way in which this formula has been derived obviously implies mutual independence among the windows, i.e they do not have to overlap, otherwise the simple multiplication of the factors contained in it would not be possible.

The (5) represents the answer to our problem: it replaces the Poisson distribution in giving the correct description of the background statistical fluctuations to be expected in the multiple windows search algorithm. Therefore, on one hand it can be used to asses the actual significances obtained if the detection threshold is improperly set on the basis of the Poisson distribution pertaining to each single window, and on the other it allows to define the proper thresholds to impose correctly the desired significance levels for the signal detection.

| Significance Definitions | $1\sigma$: 0.1587 | $2\sigma$: 0.0228 | $3\sigma$: 0.0014 | $4\sigma$: $3.2 \times 10^{-5}$ | $5\sigma$: $2.9 \times 10^{-7}$ |
|---|---|---|---|---|---|
| Poisson: corresponding threshold count | 25 | 30 | 36 | 41 | 47 |
| W=7 resulting significance | 0.696881 | 0.143086 | 0.005612 | 0.000178 | $1.33 \times 10^{-6}$ |
| W=13 resulting significance | 0.891036 | 0.249321 | 0.010397 | 0.00033 | $2.48 \times 10^{-6}$ |
| W=25 resulting significance | 0.98592 | 0.423911 | 0.019899 | 0.000635 | $4.76 \times 10^{-6}$ |

*Table I – Actual significances in the sliding windows signal search while keeping the thresholds pertaining to the Poisson distribution*

The resulting effect of the (5) is shown in Fig. 1, where the plots are referred to the toy model described in reference [4]. In this model it is assumed a flat background spanning a conventional range of interest from 0 to 100, and a series of "null" (i.e with no signal embedded) experiments each generating over this range 500 background events. The whole range is scanned through searching windows of width equal to 4, centered from the abscissa 2 to the abscissa 98, and shifted between these limits with variable step sizes. In each window, hence, the expected background events are B=20.

| Significance Definitions | $1\sigma$: 0.1587 | $2\sigma$: 0.0228 | $3\sigma$: 0.0014 | $4\sigma$: $3.2 \times 10^{-5}$ | $5\sigma$: $2.9 \times 10^{-7}$ |
|---|---|---|---|---|---|
| Poisson: corresponding threshold count | 25 | 30 | 36 | 41 | 47 |
| W=7 corresponding threshold count | 30 | 34 | 39 | 44 | 49 |
| W=13 corresponding threshold count | 32 | 35 | 40 | 44 | 50 |
| W=25 corresponding threshold count | 33 | 36 | 40 | 45 | 51 |

*Table II – Thresholds which ensure correctly the desired significances in the sliding window signal search*

In the figure the (5) is displayed for the cases of step size 16, 8 and 4, corresponding, respectively, to a number W of 7, 13 and 25 non overlapping windows; for comparison purpose in the figure it is also plotted the original Poisson distribution: clearly the "effective" noise distributions stemming from the multiple windows search are progressively shifted toward higher values, thus weakening the significance of a claimed detection if the threshold is kept fixed to the value pertaining to the Poisson case.

To better illustrate such an occurrence, in the figure it has been added a vertical line marking on the Poisson curve the 2σ value of 30 counts, i.e. the sum of the probabilities of the counts above this threshold (including 30 itself) is less than the conventional 2σ definition of 2.28% (due to the discrete nature of the Poisson distribution the sum cannot be exactly equal to the conventional definition.). Obviously, if the same threshold is confronted with the "actual" noise distributions, we see that the probability that in our search we are fooled by noise fluctuations is much higher than the expected nominal 2σ level.

In term of counts for a given significance the situation is summarized in tables I and II. In table I in the first row there are the conventional significance definitions, in the second the numbers of counts to be set as detection threshold in order to ensure those significances for the original Poisson distribution, in the third, fourth and fifth rows the actual significances corresponding to those thresholds for the 7, 13 and 25 windows distributions. From the table it can be inferred that the actual detection significances are remarkably weakened if the thresholds are kept equal to those related to the Poisson curve.

In table II the first two rows are equal to those in table I, while in the last three rows there are the threshold counts that would ensure, in the multiple search scanning, to reach truly the desired nominal significances.

## 3. Comparison with Monte Carlo

In order to check the (5) a few Monte Carlo calculations have been performed, referred to the toy model mentioned above. Repeating a series (100000) of 500 events experiments, scanned with a step size equal to 4 (i.e. 25 windows in total), the distribution of the largest count detected in the search is evaluated and plotted in Fig. 2 together with the model function (5).

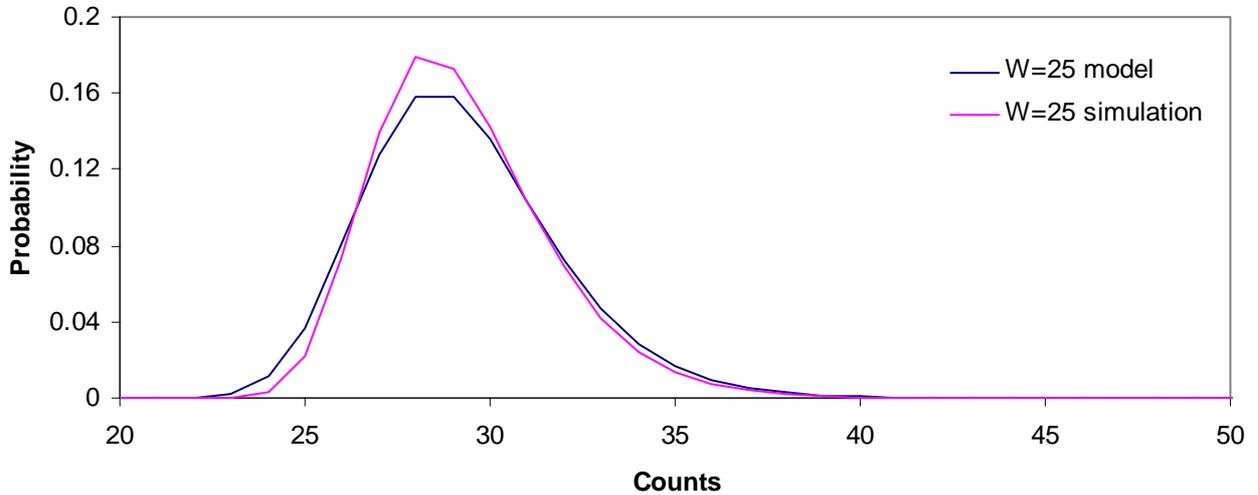

*Fig. 2. Monte Carlo – model comparison for W=25 exploring windows*

Event though the agreement is satisfactory, however clearly the Monte Carlo demonstrates a discrepancy with the model (5). The reason is that the random variables "number of counts in each search window" are neither Poisson distributed nor mutually independent, even if the windows do not overlap. Indeed, being fixed to 500 the numbers of generated events in each experiment over the range from 0 to 100, we have that the probability of an event to belong to one of the windows is $p=4/100$ (we remind 4 is the width of the window in the toy model), and that therefore the probability of $n$ counts in a windows is the binomial distribution

$$P(n) = \binom{500}{n} p^n (1-p)^{500-n} \tag{6}$$

So, the (5) should be modified at least as

$$P_{\max}(N) = \sum_{k=1}^{W} \binom{W}{k} \left( \sum_{n=0}^{N-1} \binom{500}{n} p^n (1-p)^{500-n} \right)^{W-k} \left( \binom{500}{N} p^N (1-p)^{500-N} \right)^k \tag{7}$$

However, the (7) still does not solve the discrepancy with the Monte Carlo, since the number of counts in each window cannot vary freely but it is constrained in order to keep the total equal to 500. Hence, in counting all the configurations pertaining to the probability of the maximum detected count N one should take into account such a constraint, which makes it impossible to write down a simple formula like (5) or (7). For W=25 windows, the (7) is plotted in Fig. 3 together with the simulated distribution, from which it can be inferred that, considering more appropriately the binomial instead of the Poisson distribution, it is obtained at least a good Monte Carlo-model agreement between the tails of the two curves.

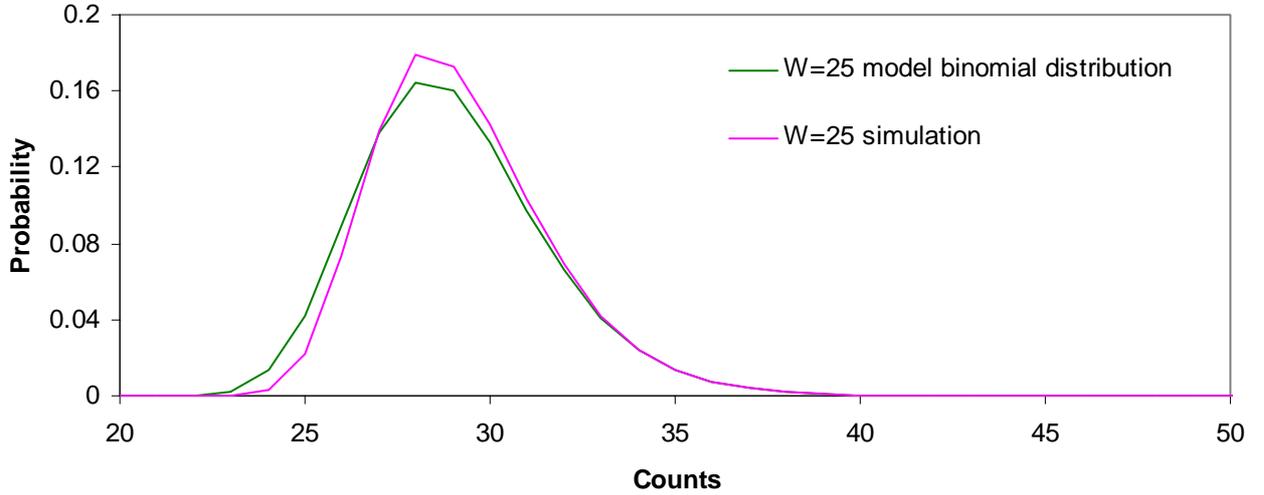

*Fig. 3. Monte Carlo – model comparison for W=25 exploring windows. The model has been modified to account for the binomial count distribution in each window*

Given the above mentioned difficulties, it is, therefore, simpler to check the Monte Carlo-model consistency by modifying the Monte Carlo so to allow truly uncorrelated Poisson windows. This can be done very simply if the toy model is modified allowing the total number of counts to fluctuate according to a Poisson distribution with mean value T=500. In this case it is well known that, being Poisson distributed the number of counts over the whole range, it is also Poisson distributed the number of counts over each sub-interval. It may be interesting to check mathematically why the distribution of the counts in each window, which is binomial when the number of events is kept fixed over the whole range, is converted into a Poisson distribution when the Poisson fluctuation is allowed for the total number of counts in the entire range. This can be done easily exploiting the concept of conditional probability; by denoting with $p(n/k)$ the probability of $n$ events in the individual window of fractional width $p$, conditioned to the total number $k$ of events in the whole interval, and with $p(k)$ the probability of $k$ events in the whole range, we have that the probability $P(n)$ of $n$ counts in the single window can be computed as

$$P(n) = \sum_{k=n}^{\infty} p(k) p(n/k) \qquad (8).$$

Since *p(n/k)* is the binomial distribution

$$p(n/k) = \binom{k}{n} p^n (1-p)^{k-n} \qquad (9)$$

and *p(k)* is the Poisson distribution

$$p(k) = \frac{e^{-T} T^k}{k!} \qquad (10)$$

then *P(n)* is given by

$$P(n) = \sum_{k=n}^{\infty} \frac{e^{-T} T^k}{k!} \binom{k}{n} p^n (1-p)^{k-n} \qquad (11)$$

that can be manipulated as follows

$$P(n) = \sum_{k=n}^{\infty} \frac{e^{-T}}{k!} \binom{k}{n} (pT)^n ((1-p)T)^{k-n} \qquad (12)$$

$$P(n) = \sum_{k=n}^{\infty} \frac{e^{-T}}{k!} \frac{k!}{n!(k-n)!} (pT)^n ((1-p)T)^{k-n} \qquad (13)$$

$$P(n) = \frac{e^{-T} (pT)^n}{n!} \sum_{k=n}^{\infty} \frac{((1-p)T)^{k-n}}{(k-n)!} \qquad (14).$$

The last sum is immediately recognized as the series expansion of *exp((1-p)T)* and hence the (14) becomes

$$P(n) = \frac{e^{-T} e^{(1-p)T} (pT)^n}{n!} \qquad (15)$$

and finally

$$P(n) = \frac{e^{-pT} (pT)^n}{n!} \qquad (16).$$

Hence, in this case the number of counts in each window truly Poisson fluctuates independently about the mean value given by the average total number of counts times the fractional width of the window itself. In such an occurrence the only condition to fulfill independency is thus that the windows do not overlap.

The model comparison with the Monte Carlo performed according to this prescription is reported in figure 4 for the step size of 4 (25 not overlapping scanning windows): now the agreement between the two curves is really very good.

**4. Overlapping Poisson windows**

Even in the full Poisson situation, the formula (5) is no more valid if the windows overlap each other, since obviously in this case the independency is lost. In such a situation one has to rely only on the Monte Carlo to predict the largest count distribution.

Let's consider for example the toy model (still with a total of 500 Poisson distributed events per experiment) with the sliding window position increased with step sizes 1 and 0.5, originating, respectively, 97 and 193 overlapping search windows. The resulting Monte Carlo distributions are plotted in Fig. 5 together with the distributions already displayed in Fig. 1.

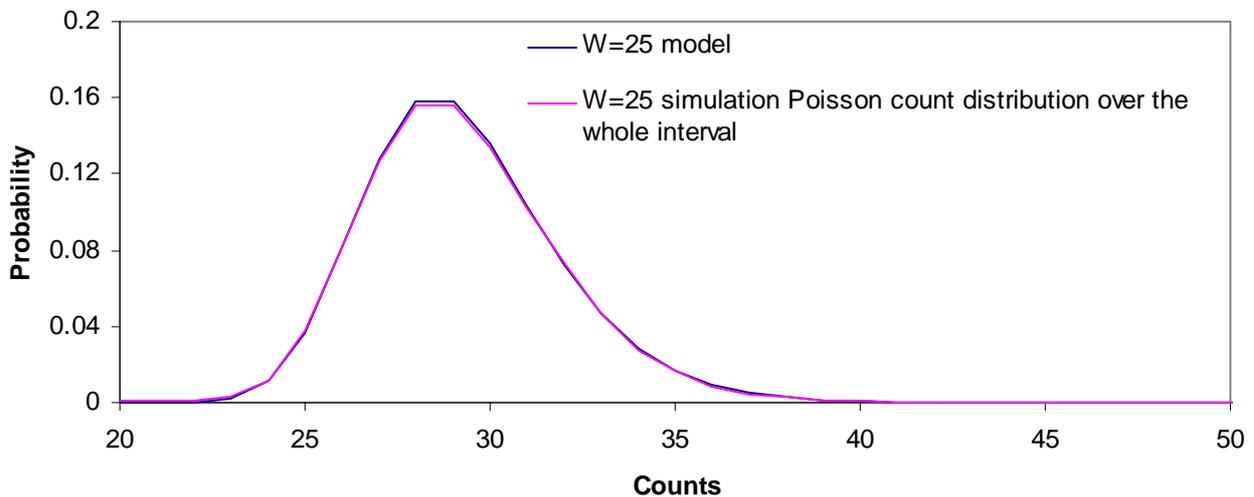

*Fig. 4. Monte Carlo – model comparison for W=25 exploring windows. The simulation has been modified to account for the Poisson count distribution over the whole interval*

As intuitive, there is a sort of saturation effect as the number of windows increases, in the sense that the strong correlation among the windows when they almost completely overlap originates only further modest shifts towards higher values of the resulting distributions of the highest detected count.

Heuristically such an occurrence can be expressively interpreted as if we were scanning an effective number of windows less than the actual number. For example in Fig. 6 there are displayed both the simulated curve for W=193 and the (5) plotted for $W_{eff}$=77 windows, which is the number that minimizes the difference between the theoretical curve and the simulated one. Even if only approximate, such a match can be referred identifying a number of effective scanned windows equal to 77. Hence, the number of truly independent scanned windows increases much slowly with respect to the actual amount of exploring windows, despite the apparent unlimited increase of the latter that could be obtained through very fine grained step sizes.

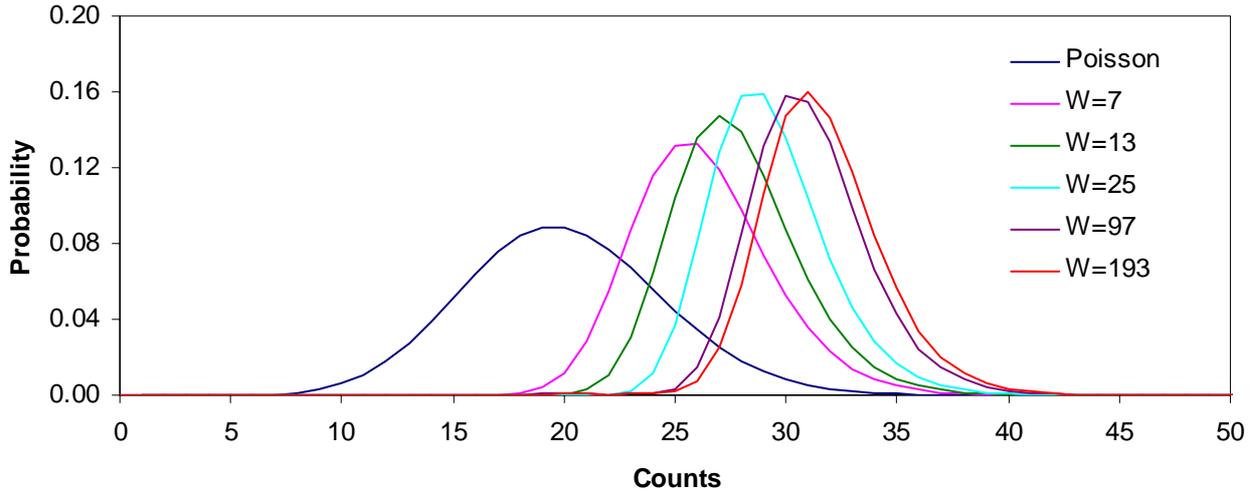

*Fig. 5. Same as Fig. 1 with the addition of the two simulated distributions for 97 and 193 overlapping exploring windows*

It is in any case worth to remark that the simulation is the only way to get the correct distribution of the largest detected count while dealing with overlapping windows.

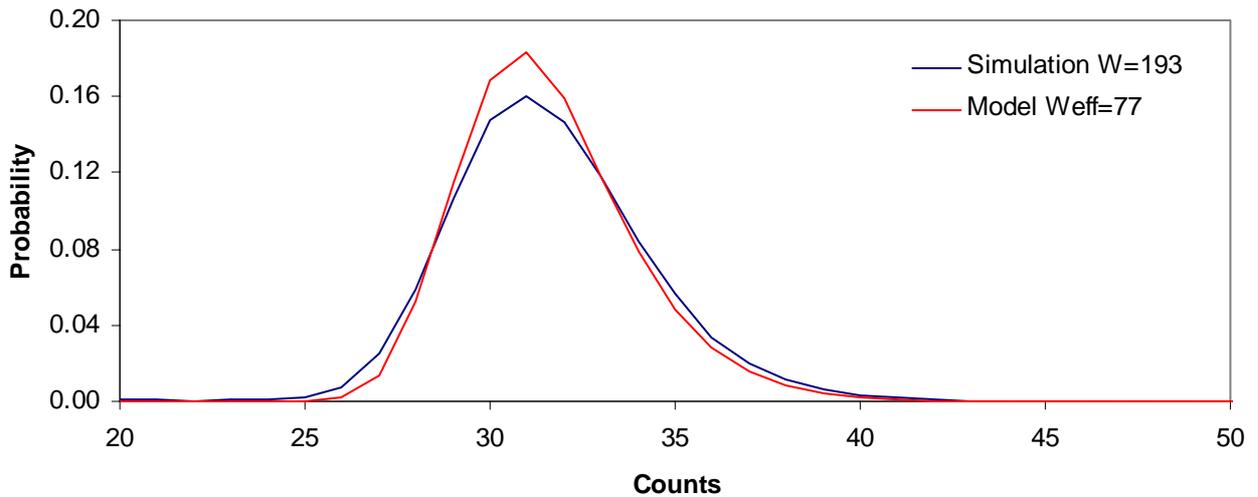

*Fig. 6. Simulated distribution for W=193 windows compared with the model for an effective number of windows equal to 77*

**5. Conclusions**

The search of a signal of unknown location is affected by noise fluctuations larger than those pertaining to a fixed location search. In this note such an effect has been demonstrated on the basis of probabilistic calculations, complementing the Monte Carlo demonstration given in [4].

Therefore, the significance of a claimed detection while scanning a broad search region for new particles must be carefully evaluated in order to avoid overestimation of the experimental discovery potential.

**Acknowledgement**

The author would like to thank Marco Rovere for help and assistance with the software programming.

**References**

[1] P.K. Sinervo, "Signal Significance in particle physics", Proceedings of the conference Advanced Statistical Techniques in Particle Physics, Durham, March 18 – 22, 2002

[2] W.A. Rolke and A.M. Lopez, "How to claim a discovery", Proceedings of PhyStat2003, SLAC, Stanford, September 8-11, 2003

[3] Y. Gao, "New Analysis Strategy to Search for New Particles at LHC", hep-ex/0310011

[4] Y. Gao, L. Lu, X. Wang, "Significance Calculation and a New Analysis Method in Searching for New Physics at the LHC", physics/0509174

[5] K. Cranmer, "Statistical challenges for searches for new physics at the LHC", physics/0511028